\documentstyle[aps,prd,psfig]{revtex}
\begin{document}
\title{Quantum radiation pressure on a moving mirror at finite temperature}
\author{L. A. S. Machado, P. A. Maia Neto\thanks{e-mail: \tt pamn@if.ufrj.br},  and C. Farina}
\address{Instituto de F\'{\i}sica, UFRJ, Caixa Postal 68528, 
21945-970 Rio de Janeiro, Brazil}
\date{\today}
\maketitle
\begin{abstract}
We compute the 
radiation pressure force on a moving mirror, in the nonrelativistic approximation, 
assuming the field to be at temperature $T.$
At high temperature, 
the force has a dissipative component proportional to the mirror velocity, which results from 
Doppler shift of the reflected thermal photons.  
In the case of a scalar field, 
the force has also a dispersive component associated to a mass correction $\Delta m.$ 
In the electromagnetic case, the separate contributions to the 
mass correction from the two polarizations cancel.
We also derive explicit results in the low temperature regime, and present numerical results for the general case. As an application, we compute the dissipation and decoherence rates
for a mirror in a harmonic potential well.

\end{abstract}

\section{Introduction}

In his seminal paper published in 1948 \cite{Casimir},  Casimir
computed the attractive force between two neutral perfectly
conducting plates due to vacuum fluctuations of the electromagnetic
field. 
The Casimir force itself is  a fluctuating quantity~\cite{barton91}, and from the 
general argument related to the fluctuation-dissipation theorem~\cite{jrqoptics} one may
expect that dissipation occurs in the case of moving boundaries. 
The energy dissipated from  macroscopic moving bodies
yields for the creation of real particles (photons in the case of the electromagnetic field)~\cite{Moore}.
Hence the vacuum radiation pressure on moving boundaries has a dissipative component that plays the 
role of a radiation reaction force. 

This effect takes place even in the case of a single moving plate, as  shown by  
Fulling and Davies~\cite{FDavies}. They treated exactly the problem of a massless scalar field in 1+1 dimensions in the presence of a  plate moving in a prescribed arbitrary way. 
However, since they employed a method based on conformal transformations, their results could not be generalized to higher
dimensions.  In order to address the case of 3+1 dimensions in the non-relativistic 
regime, a convenient perturbative method was proposed 
by Ford and Vilenkin~\cite{Ford}.
Their approach is based on the assumption that the field modification induced by the 
motion of the plate is a small perturbation, which is computed up to first
order on the displacement of the plate. 
They considered a massless scalar field, in either 1+1 or 3+1 dimensions. In the former case, 
the dissipative force is proportional to the third time derivative of the  plate's displacement, 
and corresponds to the non-relativistic limit of Fulling and Davies' result. 
For 3+1 dimensions the force on a plane mirror moving along the normal direction 
is proportional to the fifth time derivative
of the displacement. This is also the case when the electromagnetic field is considered~\cite{PA94}, 
although the proportionality factor is not simply twice the value found for the scalar case, as 
would be guessed by crude analogy with the static Casimir effect. Higher order derivatives  
appear when considering moving mirrors of finite extent~\cite{PRA93}.

A dissipative force proportional to the velocity of the mirror
(like a viscous force) would clearly violate the Lorentz
invariance of the vacuum field.
For a thermal field, on the other hand, this requirement does not hold, and 
the thermal contribution to the dissipative force turns out to be
proportional to the velocity in the case of 1+1 dimensions~\cite{JaekelReynaudPLA93}.  
The effect of thermal photons is larger than the contribution of vacuum fluctuations
to the force for temperatures larger than 
 $\hbar \omega_0/k_B$ (where $k_B$ is the Boltzmann
constant, and $\omega_0$ a typical mechanical 
frequency). 
This corresponds to temperatures in the mK range for frequencies 
in the  MHz range.
This clearly shows the importance of temperature in the dynamical Casimir effect, which would 
probably provide
the dominant contribution in any attempt to measure the force. 
Thermal effects on the generation of photons in a cavity with moving mirrors
have also been considered~\cite{thermal}.

In this paper, we 
analyze the thermal contributions  to the radiation pressure force in 3+1 dimensions, 
for both scalar and electromagnetic fields.  We take a perfectly-reflecting plane mirror
moving along the normal direction, in the non-relativistic regime. 
Our approach allows us to 
identify and distinguish between 
the field modes contributing to the dissipative component of the force
from those contributing to its dispersive component. 
We derive analytical results in the 
low and high temperature limits, 
and also compute numerically the force in the general case.
The paper is organized as follows: in the next section we take a massless scalar 
field under Dirichlet 
boundary condition. In Sec.~III, 
we consider the electromagnetic field. The results of this section are then 
applied, in Sec.~IV, to the analysis of 
dissipation and decoherence of a mirror in a potential well. 
Section \ref{conclusions} presents an interpretation of the results
in the high-temperature limit and some concluding remarks.
\section{Massless scalar field}
We choose Cartesian axis such that the plane of the mirror is parallel to the ${\cal OXY}$ plane. The mirror
is displaced  along the  
${\cal OZ}$ direction in a  prescribed, non-relativistic way. 
Hence the field $\phi$ satisfies the wave equation  and the Dirichlet boundary condition:
\begin{equation}\label{DiricheltBCphi}
\partial^2\phi=0\;\;\;\;\;\;\;\mbox{with}\;\;\;\;\;\;\;
\phi(x,y,\delta q(t),t)=0\; ,
\end{equation}
where $\delta q(t)$ denotes the position of the mirror at time $t.$ 
We assume that 
$\delta q(t)$ 
is small when compared with some characteristic field wavelength. 
We follow the perturbative approach of Ford and Vilenkin \cite{Ford} and write the field as
\begin{equation}\label{FordVilenkinmethod}
\phi=\phi^0+\delta\phi\; ,
\end{equation}
where $\phi^0$ is the solution of the corresponding static problem:
\begin{equation}\label{Eqparaphi0}
\partial^2\phi^0=0\;\;\;\;\;\;\;\;\;\;\mbox{with}\;\;\;\;\;\;\;\;\phi^0(x,y,0,t)=0\; ,
\end{equation}
and $\delta\phi$ is a small motion-induced perturbation. 
By taking the Taylor expansion around 
$z=0$ up to first order in $\delta q,$ we derive the following boundary condition:
\begin{equation}\label{deltaphiBC}
\delta\phi(x,y,0,t)=-\delta q(t)\,\partial_z\phi^0(x,y,0,t)+{\cal O}(\delta q^2)\; .
\end{equation}
We solve for $\delta\phi$ in terms of $\phi^0$ in the Fourier representation, defined
as (using capital letters for Fourier transforms)
\begin{equation}
\Phi(z,{\bf k}_{\|},\omega)=\int dt \int d^2 {\bf r}_{\parallel}\, 
e^{i\omega t}e^{-i {\bf k}_{\|}\cdot{\bf r}_{\|}}\,
\phi({\bf r}_{\|}+z \hat{\bf z},t).
\end{equation}
We find 
\begin{equation}\label{solution}
\delta\Phi(z,{\bf k}_{\|},\omega)=
- e^{i{\kappa} |z|}\, \int {d\omega_{\rm in}\over 2\pi} \delta Q(\omega-\omega_{\rm in})\partial_z
\Phi^0(0,{\bf k}_{\|},\omega_{\rm in}),
\end{equation}
where (we take $c=1$)
\[
\kappa = [(\omega+i\epsilon)^2-k_{\|}^2]^{1/2}\;\;\;\;\;\;\;\;\;\;\mbox{with}\;\;\;\;\;\;\;\;\epsilon\rightarrow 0^+\;
\]
is defined, for a given value of $k_{\|}$, as a function of $\omega$
with a branch cut along the real axis between 
$-k_{\|}$ and $k_{\|},$ so that $\kappa$ is positive for $\omega>k_{\|},$
negative for $\omega< -k_{\|},$ and equal to $i \sqrt{k_{\|}^2-\omega^2}$ otherwise. 
Then, when corresponding to a propagating field, $\delta\Phi$ propagates outwards from 
the region around the moving mirror; otherwise it corresponds to an evanescent wave. According to 
(\ref{solution}), the scattering by the moving plate generates frequency modulation: for a given 
mechanical frequency $\omega_0,$ the input field Fourier component at frequency $\omega_{\rm in}$  is 
scattered into a new frequency $\omega=\omega_{\rm in} + \omega_0.$  
Due to translational symmetry along the 
${\cal OXY}$ plane, all scattered components have the same ${\bf k}_{\|}.$ 
If $|\omega_{\rm in}+ \omega_0| <  k_{\|},$  the scattered wave  
is evanescent. 

We write the Fourier representation of the unperturbed field in the half-space 
$z > 0$ in terms of the
bosonic operators $a({\bf k})$:
\begin{equation}\label{phi0escalar}
\Phi^0(z,{\bf k}_{\|},\omega)=
i\sqrt{{16\pi^3\hbar |\omega|\over \kappa^2}}
\mbox{sin}\left(|\kappa| z\right) 
\left[a(|\kappa|\hat{\bf z}+{\bf k}_{\|})
\Theta(\omega)+a(|\kappa|\hat{\bf z}-{\bf k}_{\|})^{\dagger}
\Theta(-\omega)\right]\Theta(\kappa^2),
\end{equation}
where 
$\Theta$ denotes the step function. This equation shows explicitly the association between 
positive (negative) frequencies and annihilation (creation) operators. Moreover, 
the normal mode decomposition includes only propagating waves (since evanescent waves 
do not satisfy the required boundary condition), hence the factor $\Theta(\kappa^2).$
A similar representation may be written for the field 
in the half-space $z < 0$ in terms of independent bosonic operators $b({\bf k})$ (and their Hermitian 
conjugates).

We compute the radiation pressure force from the stress tensor component
\begin{equation}
T_{zz}={1\over 2} \left[ (\partial_x\phi)^2+(\partial_y\phi)^2-(\partial_z\phi)^2-(\partial_t\phi)^2 \right],
\end{equation} 
taken at the surface of the moving mirror. Up to first order in $\delta q,$
the force is given by 
\begin{equation}\label{force1}
\delta\!f(t) = \int dx\, dy\left(\delta T_{zz}(x,y,0^+,t)-\delta T_{zz}(x,y,0^-,t)\right)
\end{equation}
where ($\{...,...\}$ denoting the anti-commutator)
\begin{equation}\label{def_deltaT}
\delta T_{zz}(x,y,z,t)=-{1\over 2} \left\{ \partial_z\phi^0(x,y,z,t),\partial_z\delta\phi(x,y,z,t)\right\}
\end{equation}
is the motion-induced modification of the stress tensor. 
Its Fourier representation
may be computed   
from Eqs.~(\ref{solution}) and (\ref{def_deltaT}). When 
taking the average over a given field state we find
(with $\epsilon\rightarrow 0^+$)
\begin{eqnarray}\label{deltaTzz}
\langle\delta {\cal T}_{zz}(0^+,{\bf k}_{\|},\omega)\rangle_{{\;}}
&=&{i\over 2}\int{d\omega_1\over 2\pi}\int{d^2{\bf k}_1{}_{\|}\over(2\pi)^2}\, 
\left[(\omega-\omega_1+i\epsilon)^2-({\bf k}_{\|}-{\bf k}_1{}_{\|})^2\right]^{1/2}\cr\cr
&\times&
\int{d\omega_2\over 2\pi}\,\delta Q(\omega-\omega_1-\omega_2)
 \;\sigma({\bf k}_1{}_{\|}, \omega_1;
{\bf k}_{\|}-{\bf k}_1{}_{\|}, \omega_2)\; ,
\end{eqnarray}
where 
\begin{equation}\label{sigma12}
\sigma(1;2)\equiv \left\langle\left\{\partial_z\Phi^0(0^+,{\bf k}_1{}_{\|},\omega_1),
\partial_z\Phi^0(0^+,{\bf k}_2{}_{\|},\omega_2)\right\}\right\rangle
\end{equation}
is the correlation function of the unperturbed field taken at the 
${\cal OXY}$ plane. For a thermal field, 
we find, using the normal mode decomposition as given by (\ref{phi0escalar}):
\begin{equation}\label{sigma12expressao}
\sigma(1;2)  = 16\pi^3\hbar\sqrt{\omega_1^2- k_{\|}{}_1^2}\;
\Theta(\omega_1^2- k_{\|}{}_1^2)\,\delta(\omega_1+\omega_2)
\delta^{(2)}({\bf k}_1{}_{\|}+{\bf k}_2{}_{\|})
\left[1+2{\overline n}(|\omega_1|)\right]
\end{equation}
where 
\[
{\overline n}(|\omega|)=(e^{\hbar|\omega|/(k_B T)} -1)^{-1} 
\]
is the average photon number at 
temperature $T.$ 

We now analyze in detail the expression in the r.-h.-s. of Eq.~(\ref{sigma12expressao}).
The factor $1+2{\overline n}(|\omega_1|)$ originates from the general 
relation between the thermal averages of anti-commutators and commutators, 
as given by the fluctuation-dissipation theorem~\cite{kubo}. For the field operators themselves, 
the commutator is a c-number, hence the  temperature dependence comes solely from 
this factor. The factor $\delta^{(2)}({\bf k}_1{}_{\|}+{\bf k}_2{}_{\|})$
is a signature of a homogeneous field state:
it means that the correlation function for two given points on the surface of the plate depends only 
on the relative position between the points. When replaced into (\ref{deltaTzz}), it yields for 
an uniform pressure over the surface of the plate. 
Likewise, the factor $\delta(\omega_1+\omega_2)$ in Eq.~(\ref{sigma12expressao}) is a signature of 
a stationary field state. When written in the time domain, it corresponds to a correlation function 
depending only on the time difference, not on the individual times themselves. 
When replaced in (\ref{deltaTzz}), this factor singles out the mechanical 
Fourier component at the same frequency $\omega$ appearing in the argument of 
$\langle{\cal T}_{zz}(0^+,{\bf k}_{\|},\omega)\rangle.$ To compute the force from (\ref{force1}), 
we also need the motion-induced stress correction at $z=0^-,$
which is computed from (\ref{solution}) and the normal mode decomposition 
of the field in the half-space $z < 0$ in analogy with the derivation of (\ref{deltaTzz}).
Its contribution simply doubles the 
value of the net force, which we write as (we employ the superscript D to denote the results for the 
scalar field obeying Dirichlet boundary conditions)
\begin{equation}\label{def_sus}
\delta\! F^{\rm D}(\omega)= \chi^{\rm D}(\omega) \,\,\delta Q(\omega).
\end{equation} 
We replace Eq.~(\ref{sigma12expressao})  into Eq.~(\ref{deltaTzz}) and integrate over the 
${\cal OXY}$ plane to derive the susceptibility function $\chi^{\rm D}(\omega)$ ($A$ is the area of the 
mirror):
\begin{equation}\label{chiDirichlet}
\chi^{\rm D}(\omega)={i\hbar A\over \pi^2}\int d\omega_{\rm in}
\int_{0}^{|\omega_{\rm in}|} dk_{\|}\, k_{\|} \,
\biggl[(\omega+\omega_{\rm in}+ i\varepsilon)^2-k_{\|}^2\biggr]^{1/2} \, 
\sqrt{{\omega_{\rm in}}^2-k_{\|}^2}\,
\left[ {1\over 2}+{\overline n}(|\omega_{\rm in}|)\right],
\end{equation}
where we have replaced the variable of integration $\omega_2$ in (\ref{deltaTzz})
by $\omega_{\rm in},$ since it corresponds to the frequency of the 
unperturbed field $\Phi_0$ [or input  frequency, see Eq.~(\ref{solution})].

The imaginary part of the susceptibility provides, according to 
Eq.~(\ref{def_sus}), a force component in quadrature with the displacement.
If ${\rm Im}\chi^{\rm D} >0,$ this force component is in opposition of phase with
respect to the velocity of the mirror, and hence 
dissipates its mechanical energy. 
On the other hand, the real part provides the dispersive force component, which is in quadrature 
with the velocity, and does not engender any energy exchange when averaging over a 
sufficiently long time interval. 
 According to Eq.~(\ref{chiDirichlet}), 
${\rm Re}\chi^{\rm D} $ results from contributions of 
 input modes satisfying $|\omega+\omega_{\rm in}| < k_{\|},$  
i.e.  (propagating) modes that generate evanescent waves when scattered by the mechanical Fourier component 
$\omega.$ 
In Fig.~1, we represent the region for the integration in the r.-h.-s. of 
(\ref{chiDirichlet}), corresponding to the condition $|\omega_{\rm in}|\ge k_{\|} \ge 0$
(all input modes are propagating waves), in the plane $\omega_{\rm in}-k_{\|}.$ 
It is divided in four subsets, labeled $R_1$ to $R_4$~\cite{PA94}.
In this diagram, the scattering by the mechanical frequency $\omega$ 
corresponds to an horizontal  displacement, by an amount equal to $\omega,$ from 
the point of coordinates $(\omega_{\rm in},k_{\|})$ representing a given input field mode 
(for the sake of clarity we assume $\omega>0$ in the diagram). 
The contribution to the dispersive component of 
the force comes from region $R_4,$ the region that occupies the evanescent sector when shifted by $\omega,$ 
whereas $R_1$ to $R_3$ contribute to dissipation.  

\begin{figure}
\centering \leavevmode
\psfig{file=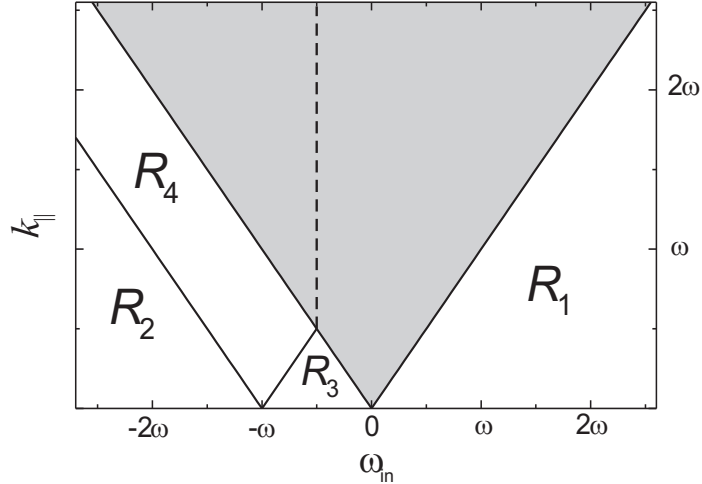,height=7.0cm,width=9.3cm}
\caption{ Diagram for the evaluation of the susceptibility, given by the 
integrals in the r.-h.-s. of (\ref{chiDirichlet}) (scalar or electromagnetic TE-polarized modes) and (\ref{chiTM}) (electromagnetic TM-polarized modes). Field modes propagating along different directions corresponding to the 
same values of $k_{\|}=|{\bf k}_{\|}|$ and frequency are represented by a single point in the $\omega_{\rm in} - k_{\|}$ plane of integration. Evanescent input modes ($|\omega_{\rm in}|<k_{\|},$ grey region) are excluded. 
Regions $R_1$ to $R_3$ yield the dissipative component of the force, whereas region $R_4,$ corresponding to input propagating waves that are scattered into evanescent waves, provides the dispersive component. At zero temperature, 
the contributions from $R_1$ and $R_2$ cancel, because the integrand in these regions is anti-symmetric with respect
to reflection around the axis at $\omega_{\rm in}=-\omega/2,$ which is indicated by a dashed line.}
\end{figure}

There are two distinct terms in the r.-h.-s. of (\ref{chiDirichlet}), both contributing to 
the dispersive and dissipative components:
one proportional to ${\overline n}(|\omega_{\rm in}|),$ 
corresponding to thermal fluctuations, 
and one independent of temperature, containing the effect of 
vacuum fluctuations (with ${\overline n}$ replaced by $1/2$). 
Accordingly, we write the susceptibility as 
\[
\chi^{\rm D} = \chi_T^{\rm D} + \chi_{\rm vac}^{\rm D}.
\]
At zero temperature, we have  by definition 
$\chi_{T = 0}^{\rm D} = 0$
and
$\chi^{\rm D}=\chi_{\rm vac}^{\rm D}.$
It is then particularly useful to consider the 
reflection around the axis 
$\omega_{\rm in}=-\omega/2$ (indicated by a dashed line in Fig.~1),
which is implemented by the transformation $\omega_{\rm in}\rightarrow -\omega -\omega_{\rm in},$
while keeping $k_{\|}$ unchanged. 
The contributions from points in region $R_2$ cancel exactly those from their 
conjugates in region $R_1$ in 
the integral in Eq.~(\ref{chiDirichlet}). As a consequence, the single contribution to  
dissipation at zero temperature comes from $R_3,$ the only bounded
region in the diagram: it corresponds to 
negative-frequency input modes that are scattered into positive-frequency propagating 
modes. We
evaluate the resulting integral to find
\begin{equation}\label{chi_vac_esc}
{\rm Im}\chi_{\rm vac}^{\rm D}(\omega)= {\hbar A \over 360 \pi^2} \omega^5,
\end{equation}
in agreement with Ref.~\cite{Ford}. 
Thus, the dissipative force exerted by the vacuum field is caused by the 
motion-induced 
mixture between positive and negative field frequencies. The discussion 
following Eq.~(\ref{phi0escalar}) indicates that this mixture is a signature 
of a Bogoliubov motion-induced transformation of creation into annihilation
operators (and vice-versa), which is clearly 
connected to the emission of particles~\cite{bogo}\cite{bogo2}. 
In fact, the dissipative force in vacuum plays the role of a 
quantum radiation reaction force, dissipating the mechanical energy at exactly
the rate required, by energy conservation, for the photon emission effect.

As for the dispersive component of the force, on the 
other hand, the integral runs over the unbounded region $R_4,$
and  the vacuum contribution diverges. After regularization, 
the dispersive force leads to renormalization of the mass of the 
mirror~\cite{Ford}, an effect analyzed in detail for 
a dielectric interface in Refs.~\cite{inertia-one-plate} and 
\cite{eberlein} and for a dispersive mirror in Ref.~\cite{barton-calogeracos}.

We now analyze the thermal contribution to the force, represented by
$\chi_T^{\rm D}.$ In contrast with the vacuum 
force, the thermal dispersive
force, as given by the integral over region $R_4,$ is finite, 
because the average photon number decreases exponentially to zero 
at high frequencies.  
In Appendix A, we show that ${\rm Re} \chi_{T}^{\rm D}(\omega)<0$
for any $\omega$ (on the other hand, the imaginary
part must be positive so as to provide dissipation). 
The ratio $\chi_{T}^{\rm D}/{\rm Im} \chi_{\rm vac}^{\rm D}$
is a function of a single parameter, $k_B T/(\hbar \omega).$ 
Numerical results for $|{\rm Re} \chi_{T}^{\rm D}|/{\rm Im} \chi_{\rm vac}^{\rm D}$
are shown in Fig.~2 (dashed line). 
The dominant contribution comes from input frequencies 
satisfying $|\omega_{\rm in}|\stackrel{<}{\scriptstyle\sim} kT/\hbar.$
When $k_B T \ll \hbar \omega,$ 
this condition implies $|\omega_{\rm in}|\ll \omega,$ and hence 
the dominant contribution to $\chi_T^{\rm D}$ comes from the close neighborhood of $\omega_{\rm in}=0$ in 
Fig.~1.
However, region $R_4$ is separated from this 
neighborhood (see fig.~1), 
and the larger contribution comes from nearly grazing field modes with frequencies close
to $\omega_{\rm in} = -\omega/2.$ The corresponding average photon number is 
${\overline n}\approx\exp(-\hbar|\omega|/2k_B T),$ hence 
${\rm Re} \chi_{T}^{\rm D}$ is exponentially small in this limit.  
In Appendix A, we derive
\begin{equation}
{\rm Re} \chi_{T\rightarrow 0}^{\rm D}(\omega) = - {1\over 2\pi} {(k_B T)^3\over \hbar^2} A \omega^2
\exp\left(-\frac{\hbar|\omega|}{2k_B T}\right).\label{re_sca_lt}
\end{equation} 

Fig.~2 suggests that $|{\rm Re} \chi_{T}^{\rm D}|$ grows according to a 
power law when $k_B T /(\hbar \omega) \gg 1.$
Neglecting terms of the order of 
$(\hbar\omega/k_B T)^2,$
we derive in Appendix A 
the following result for the 
dispersive thermal susceptibility in the high-temperature limit:
\begin{equation}\label{Tinfty}
{\rm Re} \chi_{T\rightarrow\infty}^{\rm D}(\omega) = -  {\zeta(3)\over 2\pi} {(k_B T)^3\over \hbar^2} A \omega^2,
\end{equation}
where $\zeta(3)\simeq 1.2,$  $\zeta$ denoting the Riemann zeta function~\cite{abramowitz}.
\begin{figure}
\centering \leavevmode
\psfig{file=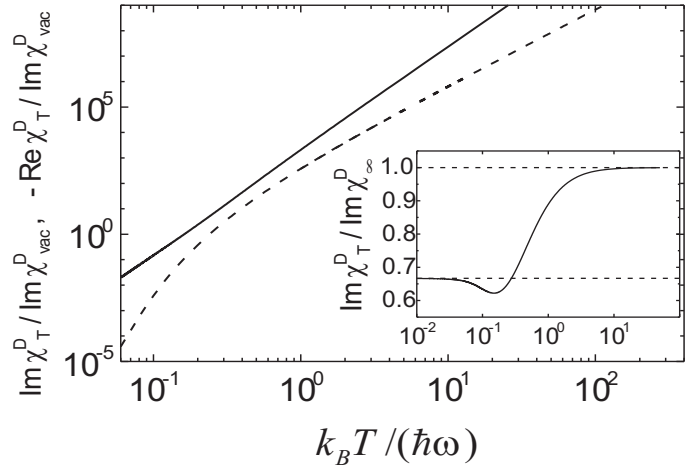,height=7.0cm,width=9.3cm}
\caption{Thermal susceptibility for the scalar field. 
The solid line represents ${\rm Im}\chi_T^{\rm D},$ whereas the dashed line represents 
$-{\rm Re}\chi_{T}^{\rm D},$ both divided by ${\rm Im}\chi_{\rm vac}^{\rm D}.$ 
In the insert, we plot ${\rm Im}\chi_T^{\rm D}$ divided by its high-temperature asymptotic value, 
showing the cross-over between the low and high temperature regimes.}
\end{figure}

The thermal dissipative susceptibility ${\rm Im}\chi_T^{\rm D}$ is computed from 
Eq.~(\ref{chiDirichlet}) in a similar way. 
We define [with $\tau \equiv k_B T/(\hbar\omega)$]
\begin{equation}\label{def_GD}
G^{\rm D}(\tau)\equiv\frac{{\rm Im} \chi_{T}^{\rm D}(\omega)}
{{\rm Im}\chi_{\rm vac}^{\rm D}(\omega)}. 
\end{equation}
In Appendix B, we derive the results
$G^{\rm D}(\tau)=24\pi^4\tau^4$ for $\tau\gg 1,$ and $G^{\rm D}(\tau)=16 \pi^4\tau^4$ for $\tau\ll 1.$ 
We plot $G^{\rm D}(\tau)$ in Fig.~2 (solid line), showing that 
the thermal contribution to dissipation becomes larger than the vacuum effect for 
$k_B T/(\hbar\omega)\stackrel{>}{\scriptstyle\sim} 0.2.$
However, the deviation from the high-temperature behavior is not visible 
in this plot. In the insert of Fig.~2, we plot the 
ratio ${\rm Im} \chi_{T}^{\rm D}(\omega)/{\rm Im} \chi_{T\rightarrow\infty}^{\rm D}(\omega),$
showing the smooth cross-over between the low and high temperature regimes. 
A similar cross-over occurs when considering  the dissipative susceptibility for the electromagnetic field,
as discussed in the next section.

\section{Electromagnetic field}\label{EMfield}

We consider the following boundary conditions for the 
electric and magnetic fields ${\bf E}'$ and ${\bf B}'$ measured in the
instantaneously co-moving Lorentz frame  $S':$
\begin{equation}\label{bc_em1}
{\hat z}\times {\bf E}'(x',y',z'=0)={\bf 0}\;\;\;\;\;\;\;{\hat z}\cdot {\bf B}'(x',y',z'=0)=0.
\end{equation}
We write
 $${\bf E}(z,{\bf k}_{\|},\omega)={\bf E}^{\rm (TE)}(z,{\bf k}_{\|},\omega)+
{\bf E}^{\rm (TM)}(z,{\bf k}_{\|},\omega),$$  
where ${\bf E}^{\rm (TE)}$ is  perpendicular to the plane defined by the vectors 
${\hat z}$ and ${\bf k}_{\|},$ whereas ${\bf E}^{\rm (TM)}$ is  parallel to this plane. 
When considering the scattering by the moving plane mirror, these two polarizations are not 
mixed, and  may be mapped into two scalar-field boundary value problems. 
For TE polarization, Eq.~(\ref{bc_em1}) yields a Dirichlet boundary condition for the 
vector potential in the laboratory frame identical to Eq.~(\ref{DiricheltBCphi}). 
The contribution of TE-polarized field modes coincides with the results found for the 
scalar field in the previous 
section: $\chi^{\rm (TE)}=\chi^{\rm D}.$

In order to 
compute the contribution from TM-polarized modes, we 
follow the approach of Ref.~\cite{PA94} and
define a new vector potential:
\begin{equation}
{\bf E}^{\rm (TM)} = \nabla\times\mbox{\boldmath$\cal A$} \; ;\;\;\;\;\;\; {\bf B}^{\rm (TM)} = 
{\partial \mbox{\boldmath$\cal A$}\over \partial t}.
\end{equation}
From Eq.~(\ref{bc_em1}), we derive a Neumann boundary condition for $\mbox{\boldmath$\cal A$}$ 
in the co-moving frame~\cite{PRA98}:
\begin{equation}\label{bcneumann}
{\partial \mbox{\boldmath$\cal A$}\over \partial z'}(x',y',z'=0)=0.
\end{equation}

From this point we proceed as in the previous section. After a lengthy
calculation, we
find for the 
contribution of TM-polarized modes
\begin{equation}\label{chiTM}
\chi^{\rm (TM)} = {i\hbar A\over \pi^2}\int d\omega_{\rm in}\int_{0}^{|\omega_{\rm in}|} dk_{\|}\, k_{\|} \,
\frac{\left[k_{\|}^2-\omega_{\rm in}(\omega+\omega_{\rm in}) \right]^2}
{\biggl[(\omega+\omega_{\rm in}+ i\varepsilon)^2-k_{\|}^2\biggr]^{1/2} \, 
\sqrt{{\omega_{\rm in}}^2-k_{\|}^2}}\,
\left[ {1\over 2}+{\overline n}(|\omega_{\rm in}|)\right].
\end{equation}
The integration region for the evaluation of $\chi^{\rm (TM)}$ is divided as shown in Fig.~1, with 
$R_4$ providing its real part, and $R_1$ to $R_3$ its imaginary part as in the scalar case.
The vacuum contribution was already discussed in detail in Ref.~\cite{PA94}. For the imaginary part, the contributions 
from regions $R_1$ and $R_2$ cancel, and $R_3$ yields a contribution larger than the TE result, so that the total 
dissipative susceptibility for the electromagnetic case is not simply twice the result of
Eq.~(\ref{chi_vac_esc}) for the scalar field:
\begin{equation}\label{EM_vac}
{\rm Im}\chi_{\rm vac}^{\rm EM} = {\rm Im}\chi_{\rm vac}^{\rm (TE)} +
{\rm Im}\chi_{\rm vac}^{\rm (TM)}= {\hbar A\over 30 \pi^2} \omega^5.
\end{equation} 
The  TM  contribution to the real part, on the other hand, cancels the $\omega^2$ (inertial) term from  TE  modes, but a $\omega^4$ divergent term remains~\cite{PA94}. 

A similar cancelation takes place when considering the thermal contribution
in the high-temperature limit.
Starting from Eq.~(\ref{chiTM}), we  show in Appendix A that 
${\rm Re} \chi_{T}^{\rm (TM)}(\omega)$ is positive for any $\omega$ (whereas
the TE contribution is negative), and 
derive
\begin{equation}\label{real_chi_TM_ht}
{\rm Re} \chi_{T\rightarrow\infty}^{\rm (TM)}(\omega) =  {\zeta(3)\over 2\pi} {(k_B T)^3\over \hbar^2} A \omega^2.
\end{equation}
Thus, the TM contribution cancels the TE contribution as given by (\ref{Tinfty}) to leading order of 
$k_B T/(\hbar\omega),$ and the resulting electromagnetic dispersive susceptibility is smaller than the 
scalar susceptibility by a factor of the order of 
$(\hbar\omega/k_B T)^2.$
For finite values of $k_B T/(\hbar\omega),$
${\rm Re} \chi_{T}^{\rm (TM)}(\omega)$  is larger than $|{\rm Re} \chi_{T}^{\rm (TE)}(\omega)|,$ 
so that ${\rm Re} \chi_{T}^{\rm EM}(\omega)$ is always positive.    
In Fig.~3, we 
plot ${\rm Re}\chi_{T}^{\rm EM}$ divided by 
$-{\rm Re} \chi_{T\rightarrow\infty}^{\rm D}(\omega)$ as 
a function of
$k_B T/(\hbar \omega),$ showing the $(k_BT/\hbar\omega)^{-2}$ behavior at high temperatures.

In the low temperature limit, on the other hand, both contributions become exponentially small.
In Appendix A, we find 
\begin{equation}\label{re_em_lt}
{\rm Re} \chi_{T\rightarrow 0}^{\rm TM}(\omega) =  
{1\over 8\pi} k_B T A \omega^4 \exp\left(-\frac{\hbar|\omega|}{2k_B T}\right).
\end{equation}
The leading contribution to both TE and TM terms originate from field modes
propagating close to grazing directions, and with frequencies close to
$-\omega/2.$ As discussed in Sec.~II, 
the resulting susceptibility is exponentially small because it is
 proportional to the average photon number at frequency $|\omega/2|.$
As compared to TE field modes [see (\ref{re_sca_lt})],  TM modes provide a 
larger contribution, by a factor of the order of $(\hbar \omega/k_B T)^2,$
so that ${\rm Re} \chi_{T\rightarrow 0}^{\rm EM}(\omega)
\approx{\rm Re} \chi_{T\rightarrow 0}^{\rm (TM)}(\omega).$ 
This is in line with the 
preference for TM polarization when 
considering propagation along a 
direction parallel to the plane of the mirror, 
since it matches
the boundary conditions in the static case. 

For the imaginary part, 
we define, in analogy with (\ref{def_GD}),
\begin{equation}\label{def_GEM}
G^{\rm EM}(\tau)\equiv\frac{{\rm Im} \chi_{T}^{\rm EM}(\omega)}
{{\rm Im}\chi_{\rm vac}^{\rm D}(\omega)}. 
\end{equation}
As discussed in Appendix B,
TM and TE modes provide identical contributions in the 
high temperature limit,  
hence ${\rm Im}\chi_T^{\rm EM}$ is simply twice the value for the scalar 
field in this limit: $G^{\rm EM}(\tau\gg 1)\approx 48 \pi^4 \tau^4.$
The TM contribution in the low temperature limit is also analyzed in Appendix B. 
We find $G^{\rm EM}(\tau\ll 1)\approx 64 \pi^4 \tau^4.$
Numerical results are presented in the next section, where
we analyze 
dissipation and decoherence in a harmonic 
potential well.

\begin{figure}
\centering \leavevmode
\psfig{file=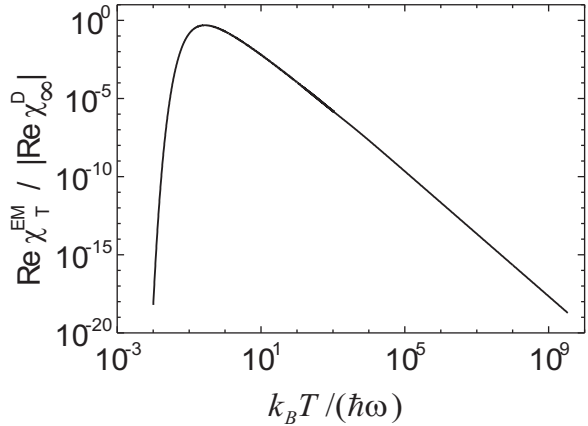,height=7.0cm,width=8.0cm}
\caption{Dispersive thermal susceptibility for the electromagnetic field. We plot
${\rm Re}\chi_T^{\rm EM}/|{\rm Re}\chi_{T\rightarrow\infty}^{\rm D}|$ 
 as a function of $k_B T/(\hbar\omega).$}
\end{figure}

\section{Dissipation and decoherence}

In this section, we consider the effect of the quantum radiation pressure 
force on the motion of the mirror. We start with a classical description of the 
position of the mirror, and analyze the damping of the mirror's oscillation 
in a harmonic potential well (frequency $\omega_0$). Later in this section we also 
consider the quantum dynamics of the mirror, in order to derive the decoherence 
rate induced by  radiation pressure. The limiting cases of zero and 
high temperatures were considered in Refs.~\cite{PRL2000} and \cite{decoherence},
respectively.
The results of the previous section allow us to take arbitrary values of 
temperature. 
  
The equation of motion reads 
\begin{equation}\label{integro-diff}
m \delta \ddot{q}(t)= - m \omega_0^2 \delta q(t) + \int dt'{\tilde\chi}(t')\delta q (t-t'),
\end{equation}
where  ${\tilde\chi}(t)$ 
denotes the inverse
Fourier transform of $\chi(\omega),$
with $\chi(\omega)=\chi^{\rm D}(\omega)$
for the scalar field model and 
$\chi(\omega)=\chi^{\rm EM}(\omega)$
for the electromagnetic case. 
Its solutions are of the form $\delta q(t) = e^{-i\Omega t} \delta q_0,$ where 
the complex constant $\Omega$ is found by replacing this function into
(\ref{integro-diff}):
\begin{equation}\label{Omega}
-m\Omega^2=-m\omega_0^2 + \chi(\Omega).
\end{equation}
We include the dispersive effect of the vacuum field through a renormalization of the bare mass and frequency of oscillation, so that 
$m$ and $\omega_0$ in (\ref{integro-diff}) and (\ref{Omega}) are already 
renormalized
and $\chi$ in (\ref{Omega}) 
is replaced by 
${\rm Re}\chi_T+i({\rm Im}\chi_{\rm vac}+{\rm Im}\chi_T).$
This allows us to 
assume that the effect generated by 
the radiation pressure
force is a small perturbation of the free oscillations at frequency $\omega_0$
when computing the roots of Eq.~(\ref{Omega}).
Hence they are of the form
$\Omega=\pm (\omega_0+\delta\omega_T)-i\Gamma,$ 
with $\Gamma, |\delta\omega_T| \ll \omega_0.$ 
$\Gamma$ is the damping 
rate, and 
$\delta\omega_T$ is the frequency shift induced by 
the coupling with thermal photons. Using that 
${\rm Re}\chi_T$  (${\rm Im}\chi$) is an even (odd) function
of $\Omega,$ we find  
from (\ref{Omega})
\begin{equation}\label{frequency_shift}
\delta\omega_T = -\frac{{\rm Re}\chi_T(\omega_0)}{2 m \omega_0}
\end{equation}
and
\begin{equation}\label{gamma_1}
\Gamma = \frac{{\rm Im}\chi(\omega_0)}{2 m \omega_0}.
\end{equation}

The case of a mass correction $\Delta m$ provides a trivial 
application of (\ref{frequency_shift}) (this is the case for the 3D scalar field for frequencies 
much smaller than $k_BT/\hbar,$ but not for the electromagnetic field): 
when the dispersive susceptibility is of the form 
${\rm Re}\chi_T(\omega)=\Delta m \;\omega^2,$ (\ref{frequency_shift}) yields 
$\delta\omega_T/\omega_0=-\Delta m/(2m),$ which is just the first order expansion of the 
exact result $(\omega_0+\delta\omega_T)/\omega_0=\sqrt{m/(m+\Delta m)}.$ 

As discussed in Sec.~II, ${\rm Im}\chi(\omega_0)$ is always positive since 
energy is taken from (and not given to) the mirror; hence $\Gamma,$ 
as given by (\ref{gamma_1}),
is positive, which
corresponds to exponential decay as required.

Of particular interest is the application of (\ref{gamma_1}) to the 
electromagnetic case, which we 
analyze in the rest of this section.  
At zero temperature, Eqs.~(\ref{EM_vac}) and (\ref{gamma_1}) yield (here and 
in the next section we re-introduce the speed of light $c$)
\begin{equation}\label{gamma_vac}
\Gamma_{\rm vac}=\frac{1}{30\pi^2}\,\frac{\hbar \omega_0}{2m c^2}\,\frac{A \omega_0^2}{c^2} \;\omega_0.
\end{equation}
Hence $\Gamma_{\rm vac}$ is proportional to the ratio between the zero point and the rest mass energies 
of the mirror. Despite of the geometrical factor $A \omega_0^2/c^2,$ 
representing the squared ratio between the transverse size of the mirror and the typical 
vacuum field wavelength (for frequencies in the 
GHz range $c/\omega_0$ is in the centimeter range), 
we have $\Gamma_{\rm vac}\ll \omega_0,$ as required for
consistency of the derivation.

In the high-temperature limit, 
Eq.~(\ref{gamma_1}) yields 
\begin{equation}\label{gamma_HT}
\Gamma_{T\rightarrow\infty} = \frac{\pi^2}{15}
\frac{k_B T}{m c^2} 
 \frac{A\left(\frac{k_B T}{\hbar}\right)^3}{c^2}, 
\end{equation}
in agreement with Ref.~\cite{decoherence}.  Except for a numerical factor, 
(\ref{gamma_HT}) differs from (\ref{gamma_vac}) by the replacement of the 
zero point energy $\hbar \omega_0/2$ by $k_B T$ and the frequency scale $\omega_0$ by 
$k_B T/\hbar.$ 
In Fig.~4, we plot 
\[
\frac{\Gamma}{\Gamma_{\rm vac}}=1+\frac{{\rm Im}\chi_T^{\rm EM}}{{\rm Im}\chi_{\rm vac}^{\rm EM}}=
1+\frac{1}{12}\;G^{\rm EM}\left(\frac{k_B T}{\hbar\omega_0}\right)
\]
as a function of $k_B T/(\hbar\omega)$
(solid line), showing the $T^4$  behavior corresponding to the high-temperature approximation 
as given by Eq.~(\ref{gamma_HT}), which is only $4\%$ larger than the numerical exact value at 
$k_B T/(\hbar\omega_0)=1.$

In order to analyze the effect of decoherence, we now consider 
the quantum dynamics of the mirror in the potential well. 
We assume that the mirror state is initially a coherent superposition of two 
coherent states: $|\psi\rangle =\left(|\alpha_0\rangle + |-\alpha_0\rangle\right)/2,$ with
\[
\alpha_0 = \frac{1}{4} \frac{\Delta Z}{\Delta Z_0}\gg 1,
\]
where $\Delta Z$ represents the distance between the wavepacket components at $t=0,$
and $\Delta Z_0=\sqrt{\hbar/2m\omega_0}$ is the uncertainty of position of the ground state.
 
The free evolution preserves the form (and the coherence) of the initial state, with
$\alpha_0$ replaced by $\alpha(t)=\alpha_0 \exp(-i \omega_0 t).$ The very weak radiation pressure coupling 
with the field introduces two new time scales, both
assumed to be much larger than the free evolution time scale $2\pi/\omega_0.$ 
The  time  needed to reach thermal equilibrium is the largest time scale. 
At zero temperature, it corresponds to 
$1/(2\Gamma),$
where $\Gamma$ is the amplitude damping rate discussed 
before in the classical 
theory. In fact, in this particular case, the equilibrium state is of
course the ground state of the harmonic oscillator. During the decay process,
the energy quantity $(2\alpha_0^2-1/2) \hbar\omega_0$ is dissipated 
by the dynamical Casimir effect, with 
the emission of about $2\alpha_0^2$ pairs of photons, since the energy 
contained in 
each pair 
is $\hbar\omega_0$~\cite{bogo2}-\cite{eberlein}\cite{lambrecht_prl}. 

\begin{figure}
\centering \leavevmode
\psfig{file=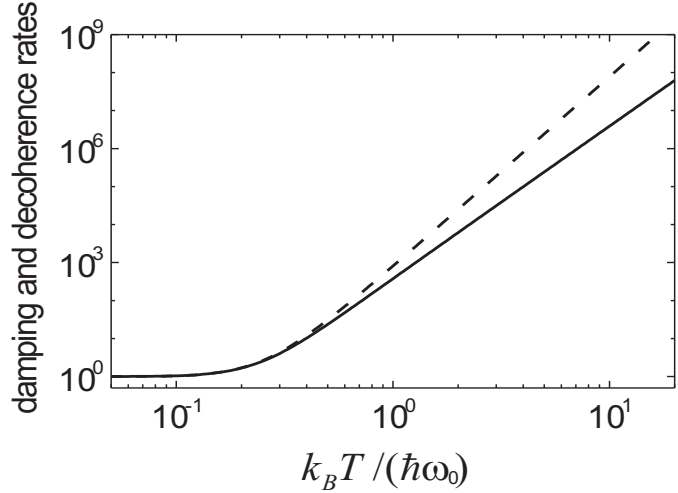,height=7.0cm,width=9.3cm}
\caption{Damping and decoherence rates for a mirror in a harmonic potential well. We plot 
$\Gamma/\Gamma_{\rm vac}$  (solid line) and 
$\Gamma^{\rm dec}/\Gamma^{\rm dec}_{\rm vac}$ (dashed line) as functions of $k_B T/(\hbar\omega_0).$
Results for the damping and decoherence rates at zero temperature $\Gamma_{\rm vac}$
and $\Gamma^{\rm dec}_{\rm vac}$ are discussed in the text; note that 
$\Gamma_{\rm vac} \ll \Gamma^{\rm dec}_{\rm vac}.$
}
\end{figure}

The second time scale corresponds to the 
process of decoherence.
The state decays into the incoherent statistical mixture 
$\rho=\left(|\alpha\rangle\langle\alpha| + |-\alpha\rangle\langle -\alpha|\right)/2$
much before 
reaching thermal equilibrium. 
The decoherence time is much
shorter than the damping time because
a single photon pair already contains `which way' information sufficient 
to destroy the possibility of interference between the two components. 
Since the time scale corresponding to the emission of 
$2\alpha_0^2$ pairs is $1/(2\Gamma),$ the decoherence time at zero temperature is
$1/(4\alpha_0^2 \Gamma)= 4(\Delta Z_0/\Delta Z)^2  (1/\Gamma)\ll 1/\Gamma.$
The decoherence rate $\Gamma^{\rm dec}$ is defined as the inverse of the corresponding time scale.  
Assuming $\Gamma^{\rm dec}\ll \omega_0,$ Ref.~\cite{decoherence}
derives a relation between decoherence and damping rates valid 
for arbitrary values of temperature:
\begin{equation}\label{descoerencia-dissipacao}
\Gamma^{\rm dec}=\frac{1}{4}\coth\left(\frac{\hbar\omega_0}{2k_B T}\right)\,\left(\frac{\Delta Z}{\Delta Z_0}
\right)^2 \,\Gamma.
\end{equation}
Formally, the factor $\coth\left(\hbar\omega_0/2k_B T\right)$ originates 
from the fluctuation-dissipation theorem, since decoherence may be described  
as a process of  diffusion in phase space, which washes out 
the interference oscillations characterizing the coherence of the superposition
state~\cite{foot_2}. 
At high temperatures, this factor is approximately $2k_B T/(\hbar \omega).$ Jointly with the 
high-temperature behavior of $\Gamma,$ it results in 
$T^5$ dependence. This power law, also 
obtained in Ref.~\cite{joos} in the case of a free particle,    
is  visible in Fig.~4 (dashed line), where 
we plot 
$\Gamma^{\rm dec}/\Gamma^{\rm dec}_{\rm vac}=\coth(\frac{\hbar\omega_0}{2k_B T})\,\Gamma/\Gamma_{\rm vac}$ 
as a function of $k_B T /(\hbar\omega_0).$  
Both $\Gamma$ and $\Gamma^{\rm dec}$ are already one order of magnitude larger 
than the vacuum values for $k_B T/(\hbar\omega_0)\simeq 0.4,$ showing that the thermal effect dominates
even for moderate values of temperature. 

According to (\ref{descoerencia-dissipacao}), 
decoherence is faster the larger is the separation $\Delta Z$ between the wave packets. 
For large separations, `which way' information is more readily available since 
the two state components are better resolved, 
as compared to their widths in phase space. 
At zero temperature, 
zero-point fluctuations define the corresponding scale of length 
$\Delta Z_0.$  
In the high-temperature limit, on the other hand, 
the additional factor $2k_B T/(\hbar\omega_0)$ in 
(\ref{descoerencia-dissipacao})
yields $\Gamma^{\rm dec}/\Gamma\approx (\Delta Z/\lambda_T)^2/2,$
the thermal de Broglie wavelength $\lambda_T=\hbar/\sqrt{2 m k_B T}$ 
replacing $\Delta Z_0$ as the reference of length scale. 

In conclusion, coherent superpositions of states very far apart in phase space are
extremely unstable, since they decay into the corresponding statistical mixtures in a 
very short time. In the opposite extreme, coherent states are the most stable ones, in the 
sense that they generate the least entropy when interacting with the field~\cite{decoherence}. They 
correspond to the `pointer states', which play an important role in the 
classical limit of quantum mechanics~\cite{deco_ref}. 
Here we have discussed a particular physical mechanism
of decoherence,  which illustrates some of its  general properties.  The 
corresponding orders of magnitude are discussed in the end of the next section.

\section{Discussion and final remarks}\label{conclusions}

The temperature of the field defines the frequency scale $k_B T/\hbar.$ The motion is slow when the 
mechanical frequencies satisfy $\omega \ll  k_B T/\hbar.$ Hence the quasi--static regime 
corresponds to the high--temperature limit. We summarize below the results obtained for this
regime, now taking the time domain.  

For the scalar field, we have found
\begin{equation}\label{viscous_force}
\delta\!f^{\rm D}(t) = - {\pi^2\over 15} {(k_B T)^4\over \hbar^3 c^4} A \,\delta \dot{q}(t) +  
{\zeta(3)\over 2\pi} {(k_B T)^3\over \hbar^2 c^4} A \,\delta \ddot{q}(t).
\end{equation}
The dissipative force results, in general, from input modes corresponding to regions 
$R_1,$ $R_2$ and $R_3$ in Fig.~1. $R_3$
corresponds to the input modes involved in the process of photon creation.
At zero temperature, only $R_3$ contributes, whereas 
in the quasi-static (high--temperature)
limit $R_1$ and $R_2$ provide the dominant contribution as shown in Appendix B.
Thus, the dissipative component in (\ref{viscous_force}) (first
term in its r.-h.-s.)
originates from scattering of thermal photons, rather than from creation of photons.
It is of the form~\cite{braginsky} $-2 P_T/c^2\;\delta \dot{q}(t),$ where $P_T,$
representing the total power incident on (both sides of) the mirror,
is proportional to $T^4$ as predicted by the Stefan-Boltzmann law. 
This viscous force results from Doppler shifting the 
reflected thermal photons~\cite{JaekelReynaudPLA93}.

The dispersive component also appearing in (\ref{viscous_force}) 
corresponds to  a temperature-dependent, finite,
and very small mass correction
\begin{equation}\label{res_mass}
\Delta m = -{\zeta(3)\over 2\pi} {(k_B T)^3\over \hbar^2 c^4} A. 
\end{equation}
For $T=300 K,$ we find $\Delta m /A \approx -1.5\times 10^{-29} {\rm g/cm^2}.$
$\Delta m$ results from the contribution of input propagating modes that are 
scattered into evanescent waves (region $R_4$ in Fig.~1). 
It corrects the (experimentally known) zero-temperature mass of the 
mirror, which already contains the mass renormalization generated by vacuum 
fluctuations~\cite{inertia-one-plate}-\cite{barton-calogeracos}. Even at zero temperature, the 
mass is modified when a second plate is present~\cite{JRmass}\cite{Ddimensions},  or if its surface is corrugated
\cite{kardar}. In all three cases, the mass correction is a finite measurable quantity that
depends on some control parameter: temperature, distance between the plates, or amplitude of 
corrugation. 
 
For the electromagnetic field, 
the viscous force is twice as large as in the scalar case, the two polarizations providing 
identical contributions:
\begin{equation}\label{final_EM}
\delta\!f^{\rm EM}(t) = - {2\pi^2\over 15} {(k_B T)^4\over \hbar^3 c^4} A \,\delta \dot{q}(t).
\end{equation}
Eq.~(\ref{final_EM}) agrees with Ref.~\cite{decoherence}, where 
the force is derived by considering the Doppler shift of the thermal photons. 
For $T=300 {\rm K},$ the pressure is $4.1\times 10^{-14} {\rm N/m^2}$ for a velocity of 
$1{\rm m/s}.$  

For the dispersive force, on the other hand, 
the contributions from 
the two field polarizations cancel, as discussed in Sec.~III,
and the quasi--static force as given by (\ref{final_EM})
does not contain a term proportional
to the second-order time derivative.
Thus, in contrast to the scalar case, 
the mass correction vanishes
in the electromagnetic case (it also vanishes 
for a scalar field in one dimension~\cite{JaekelReynaudPLA93}).
This difference between the two models 
is better understood by considering the thermodynamics of the field 
at temperature $T$ in the presence of a {\it static} mirror.
In Appendix C, we compute the Helmholtz free energy 
\begin{equation}\label{def_F}
{\cal F} = - k_B T \ln Z,
\end{equation}
where $Z$ is the partition function, for both models.
For the scalar field, the free energy in a volume $V$ is 
${\cal F}^{\rm D}={\cal E}^{\rm D}_{\rm vac}+{\cal F}^{\rm D}_T,$
where ${\cal E}^{\rm D}_{\rm vac}$
is the zero-point energy, and   
\begin{equation}\label{F_scalar}
{\cal F}^{\rm D}_T = -\frac{\pi^2}{90}\frac{(k_BT)^4}{(\hbar c)^3}V + 
\frac{\zeta(3)}{4\pi}\frac{(k_BT)^3}{(\hbar c)^2} A.
\end{equation}
The first term in (\ref{F_scalar})
is the free-space thermal free energy, featuring the characteristic $T^4$ power law. For us, the interesting 
term is the second one. It is an effect of the Dirichlet 
boundary condition at the mirror's surface, which 
eliminates field modes propagating along directions parallel to the mirror. 
In fact, this term is not present in the 
electromagnetic case:
\begin{equation}\label{F_em}
{\cal F}^{\rm EM}_T = -\frac{\pi^2}{45}\frac{(k_BT)^4}{(\hbar c)^3}V,
\end{equation}
because of the contribution from TM-polarized grazing field modes
(whereas the free-space term is twice the scalar result due to the equal 
contributions from the two orthogonal polarizations). 

From (\ref{F_scalar}) and (\ref{F_em}) we may also compute the entropy 
$S=-\partial {\cal F}/\partial T=-\partial {\cal F}_T/\partial T$ and 
the internal energy $U={\cal F}+T S.$ For the scalar field, we find
$U^{\rm D} = {\cal E}^{\rm D}_{\rm vac} + U^{\rm D}_T,$ with 
\begin{equation}
U^{\rm D}_T = \frac{\pi^2}{30}\frac{(k_BT)^4}{(\hbar c)^3}V + \Delta U_T,
\end{equation}
and $\Delta U_T=\Delta m \;c^2.$ 
$\Delta U_T$ represents the difference between 
the thermal energies in the presence of the static mirror and in free space.
As discussed in Appendix D, where 
we present an alternative, `local'
derivation of $U^{\rm D}_T,$ 
the energy density is decreased 
significantly (with respect to
its free-space value) over a 
distance from the mirror 
of the order of $\hbar c/k_B T.$

For the electromagnetic field, on the other hand,
both ${\cal F}^{\rm EM}_T$ and $U^{\rm EM}_T$ are not modified by the 
mirror according to (\ref{F_em})
(this may also be inferred from the result of 
Ref.~\cite{brown-maclay} for 
a two-plates configuration by taking the limit of large 
separation), and consistently
the  mass correction vanishes in this case.

In the low temperature (or high mechanical frequency) limit, thermal photons 
provide a small correction of the vacuum dissipative force, of the order 
of $(k_B T/\hbar\omega)^4\ll 1,$ whereas the thermal dispersive correction is exponentially small, 
for both scalar and electromagnetic cases. We have also presented numerical results 
for arbitrary values of temperature, 
allowing us to discuss damping and decoherence of a mirror in a 
harmonic potential well in this general case. 
For both effects, thermal fluctuations dominate over zero-point 
fluctuations for temperatures above $0.3 \hbar \omega_0/k_B,$
and the high temperature approximation provides  accurate  values 
for temperatures above $\hbar \omega_0/k_B.$
Although the effect of damping of energy is usually negligible, 
thermal radiation pressure efficiently  
destroys the coherence of a quantum superposition state.
For $T = 300 {\rm K}$ and $A = 1 {\rm cm^2},$ 
the decoherence time is $1/\Gamma^{\rm dec}=1.3\times 10^{-12} {\rm s}$ when the distance between the 
wave packet components is only $\Delta  Z = 1 {\rm nm}$ \cite{foot_3}.

\acknowledgements
P. A. M. N. thanks D. Dalvit,  A. Lambrecht and S. Reynaud  
for discussions, and
PRONEX, FAPERJ and the
Millennium Institute of Quantum
Information           
for financial support.
C. F. and P. A. M. N. thank CNPq for partial financial support.

\appendix

\section{Dispersive thermal susceptibility}

As discussed in Sec.~II, the dispersive thermal susceptibility 
${\rm Re} \chi^{\rm D}_T(\omega)$ is given by the term proportional to 
${\overline n}$ in
Eq.~(\ref{chiDirichlet}), to be integrated over the region $R_4$ 
in Fig.~1 (for the sake of clarity we assume $\omega>0$ in the derivation so as to be able to refer to Fig.~1):
\begin{equation}\label{real_chi_1}
{\rm Re}\, \chi^{\rm D}_T(\omega)=-\frac{\hbar A}{\pi^2}\int\int_{R_4} d\omega_{\rm in} dk_{\|} k_{\|} \sqrt{k_{\|}^2-(\omega+\omega_{\rm in})^2}
\sqrt{\omega_{\rm in}^2-k_{\|}^2}\; {\overline n}(-\omega_{\rm in})
\end{equation}
Eq.~(\ref{real_chi_1}) shows that ${\rm Re}\, \chi^{\rm D}_T(\omega)<0$ for any $\omega.$
We change to the variables
\begin{equation}
u=\frac{\hbar(k_{\|}-\omega_{\rm in})}{k_B T},\,\,\,v=-\frac{\hbar(k_{\|}+\omega_{\rm in})}{k_B T}.
\end{equation}
The Jacobian for this transformation yields
\begin{equation}\label{Jacob}
|J| \equiv \left|\frac{\partial(u,v)}{\partial(k_{\|},\omega_{\rm in})}\right|= 2 \left(\frac{\hbar}{k_B T}\right)^2.
\end{equation}
From (\ref{real_chi_1})
and (\ref{Jacob}) we find 
\begin{equation}\label{real_integral}
{\rm Re}\, \chi^{\rm D}_T(\omega)=-\frac{\hbar A}{4\pi^2}\left(\frac{k_B T}{\hbar}\right)^5\int_0^{\Omega}dv
\int_{\Omega}^{\infty} du\, (u-v) \frac{\sqrt{uv (u-\Omega)(\Omega-v)}}{e^{\frac{u+v}{2}}-1},
\end{equation}
where $\Omega=\hbar \omega/(k_B T).$
We use (\ref{real_integral}) to calculate ${\rm Re}\, \chi^{\rm D}_T(\omega)$ numerically
(see figure 2).
Because of the exponential factor, the dominant contribution 
comes from values $u\stackrel{\scriptstyle <}{\scriptstyle\sim} 1.$ 
In the high-temperature approximation, $\Omega\ll 1,$ we  have 
$v\le \Omega \ll u.$ Hence we may neglect $\Omega^2$ and $\Omega v$ 
inside the root and $v$ in the exponential in
(\ref{real_integral}),  and replace the lower limit of integration over 
$u$ by zero. We get 
\begin{equation}\label{real_estatico}
{\rm Re}\, \chi^{\rm D}_T(\omega) = 
-\frac{\hbar A}{4\pi^2}\left(\frac{k_B T}{\hbar}\right)^5
\int_0^{\Omega}dv\sqrt{v(\Omega-v)}
\int_{0}^{\infty} du\,  \frac{u^2}{e^{\frac{u}{2}}-1}\left[1+{\cal O}(\Omega^2)\right]
\end{equation}
The remaining integrals in (\ref{real_estatico}) give  
$2\pi\zeta(3)\Omega^2,$
yielding the  result of (\ref{Tinfty}).

In the low--temperature limit, $\Omega\gg 1,$ we may approximate the average photon number
 as follows: $$ \left\{\exp\left[(u+v)/2\right]-1\right\}^{-1}\approx \exp\left[-(u+v)/2\right],$$ 
because $(u+v)/2\ge \Omega/2$ in (\ref{real_integral}). 
We change to new variables $s$ and $t$ with $v/\Omega=s^2$ and 
$u/\Omega=1+t^2$ and integrate over $t$  by parts to find 
\begin{equation}\label{re_chi_lt}
{\rm Re}\, \chi^{\rm D}_T(\omega) \approx
-\frac{\hbar A}{4\pi^2}\left(\frac{k_B T}{\hbar}\right)^5
\Omega^4 e^{-\Omega/2}
\int_{-1}^{1} ds\, s^2\sqrt{1-s^2} \, e^{-\Omega s^2/2}
\int_{-\infty}^{\infty} dt \frac{\partial}{\partial t}\left[t\sqrt{1+t^2}(1+t^2-s^2)\right]
e^{-\Omega t^2/2}.
\end{equation}
The integral over $t$ gives $\sqrt{2\pi/\Omega}\, (1-s^2)$ by the method of 
steepest descent,
the saddle point  $t=0$ corresponding to 
the condition $k_{\|}-\omega_{\rm in}=\omega.$
The resulting integral over $s$ is also calculated by 
first integrating by parts, and then using the steepest 
descent method. The saddle point is at $s=0,$
corresponding to $\omega_{\rm in} = - k_{\|}.$
Thus, the dominant contribution in (\ref{real_chi_1}) comes from the close neighborhood of 
$k_{\|}=-\omega_{\rm in}=\omega/2,$ which corresponds to nearly
grazing modes, and the final expression is given 
by (\ref{re_sca_lt}).

As discussed in Sec.~III, the contribution of TE-polarized modes to the 
electromagnetic susceptibility turns out to be identical to $\chi^{\rm D}.$ The TM 
contribution is given by the r.-h.-s. of (\ref{chiTM}),
its real $T-$ dependent part yielding
\begin{equation}\label{real_chi_TM}
{\rm Re}\, \chi^{\rm (TM)}_T(\omega)= \frac{\hbar A}{4\pi^2}\left(\frac{k_B T}{\hbar}\right)^5\int_0^{\Omega}dv
\int_{\Omega}^{\infty} du\, \frac{u-v}{e^{\frac{u+v}{2}}-1}
 \frac{\left[uv-\frac{\Omega}{2}\left(u+v\right)\right]^2}{\sqrt{uv (u-\Omega)(\Omega-v)}},
\end{equation} 
It is clear from this equation that ${\rm Re}\, \chi^{\rm (TM)}_T(\omega)$ is positive for
any frequency $\omega.$
Using the same methods employed 
in the scalar case, we derive from this equation
the limiting cases corresponding to high and low temperatures.

In the high-temperature limit, we derive from (\ref{real_chi_TM})
\begin{equation}\label{real_chi_TM_aux_ht}
{\rm Re}\, \chi^{\rm (TM)}_T(\omega) = 
\frac{\hbar A}{4\pi^2}\left(\frac{k_B T}{\hbar}\right)^5
\int_0^{\Omega}dv\frac{\left(\frac{\Omega}{2}-v\right)^2}{\sqrt{v(\Omega-v)}}
\int_{0}^{\infty} du\,  \frac{u^2}{e^{\frac{u}{2}}-1}\left[1+{\cal O}(\Omega^2)\right].
\end{equation}
The integrals above are readily calculated, resulting in (\ref{real_chi_TM_ht}).

In the low temperature limit, we change to the variables 
$s$ and $t$ employed in the derivation of the scalar susceptibility, and derive 
from (\ref{real_chi_TM})  
\begin{equation}\label{re_chi_lt}
{\rm Re}\, \chi^{\rm (TM)}_T(\omega) \approx
\frac{\hbar A}{4\pi^2}\left(\frac{k_B T}{\hbar}\right)^5
\Omega^5 e^{-\Omega/2}
\int_{-1}^{1} ds \frac{e^{-\Omega s^2/2}}{\sqrt{1-s^2}}
\int_{-\infty}^{\infty} dt 
\frac{1+t^2-s^2}{\sqrt{1+t^2}}
\left[(1+t^2)(s^2-1/2)-s^2/2\right]^2
e^{-\Omega t^2/2}.
\end{equation}
The integral over $t$ gives $\sqrt{2\pi/\Omega} \, (1-s^2)^3/4$ by the method of 
steepest descent (saddle point at $t=0$). Then the integral over $s$ is also computed 
with this method (saddle point at $s=0$), yielding the result of (\ref{re_em_lt}). As in the 
scalar case, the dominant contribution comes from nearly grazing waves with
frequencies close to $-\omega/2.$

\section{Dissipative thermal susceptibility}

We calculate $G^{\rm D}(\tau),$
defined in (\ref{def_GD}),
from (\ref{chiDirichlet}), taking the imaginary part of the
term proportional to 
${\overline n}.$  
We change to the dimensionless variables 
$\Omega'=\hbar\omega_{\rm in}/(k_B T)$ and 
$K=\hbar k_{\|}/(k_B T).$
For the contribution from region $R_2$ in Fig.~1, we 
also change $\Omega'$ into $-\Omega-\Omega',$
with $\Omega=\hbar\omega/(k_B T)=1/\tau.$ The joint contribution 
from $R_1$ and $R_2$ is given by
\begin{equation}\label{R1+R2}
G^{\rm D}_{(1+2)}(\tau)= 360
\tau^5(e^{\Omega}-1)\int_0^{\infty}d\Omega' \int_0^{\Omega'} dK\,
K\, \frac{\sqrt{(\Omega'^2-K^2)\left[(\Omega+\Omega')^2-K^2\right]}\,e^{\Omega'}}{(e^{\Omega'}-1)(e^{\Omega+\Omega'}-1)}. 
\end{equation}
We split region $R_3$ into two sub-regions, corresponding to the intervals 
$-\omega\le \omega_{\rm in} \le -\omega/2,$ and $-\omega/2\le \omega_{\rm in} \le 0.$ 
 We change $\Omega'$ into 
 $\Omega+\Omega'$ when integrating over the first sub-region, and into 
$-\Omega'$
for the second sub-region. We find 
\begin{equation}\label{R3}
G^{\rm D}_{(3)}(\tau)= 360
\tau^5 \int_0^{\Omega/2}d\Omega' \int_0^{\Omega'} dK\,
K\, \sqrt{(\Omega'^2-K^2)\left[(\Omega-\Omega')^2-K^2\right]}
\left(
\frac{1}{e^{\Omega'}-1}+\frac{1}{e^{\Omega-\Omega'}-1}\right). 
\end{equation}

In the high-temperature limit, $\Omega\ll 1,$ we take
$e^{\Omega}-1\approx \Omega$ and neglect $\Omega$ in the integrand in (\ref{R1+R2}).  
We find
$G^{\rm D}_{(1+2)}(\tau\rightarrow\infty)=24\pi^4\tau^4.$
For the evaluation of $G^{\rm D}_{(3)},$ we take $1/(e^{\Omega'}-1)\approx 1/\Omega'$
[and likewise for $1/(e^{\Omega-\Omega'}-1)$] in (\ref{R3}) since $\Omega'$ is bounded by
$\Omega/2.$ The final result is ${\cal O}(\tau)$ and can be neglected. This completes the derivation of the 
high-temperature limit of $G^{\rm D}(\tau).$

In the low-temperature limit, $\Omega\gg 1,$ 
we replace $1/(e^{\Omega+\Omega'}-1)$ by $e^{-\Omega-\Omega'},$ canceling
the exponential factors in the numerator in (\ref{R1+R2}). Actually, 
this is equivalent to neglect the contribution from $R_2.$
In fact, in this approximation only the close neighborhood of the origin in 
Fig.~1 contributes (as discussed in Appendix A, the dispersive component is  
exponentially small by a similar reason, since it results from region $R_4,$ which is far from the origin).
We also take 
$\sqrt{(\Omega \pm \Omega')^2 - K^2}\approx \Omega$ in (\ref{R1+R2}) and 
(\ref{R3}), and in (\ref{R3}) neglect $1/(e^{\Omega-\Omega'}-1)$  and replace the
upper bound of the integral over $\Omega'$ by infinity.           
In this approximation, the contributions from $R_1$ and $R_3$ are equal, and we find
\begin{equation}\label{T0}
G^{\rm D}_{(1+2)}(\tau\rightarrow 0) = G^{\rm D}_{(3)}(\tau\rightarrow 0)=
120\tau^4 
\int_0^{\infty} d\Omega' \frac{\Omega'^3}{e^{\Omega'}-1}.
\end{equation}
The remaining integral in (\ref{T0}) gives $\pi^4/15,$
completing the derivation of 
the low temperature limit of  $G^{\rm D}(\tau).$

In order to derive the dissipative susceptibility for the electromagnetic 
field, we need to consider the contribution of TM-polarized modes, which is
given by the $T-$ dependent imaginary part of the expression given by 
(\ref{chiTM}).
The contributions 
from $R_1$ and $R_2$ give
\begin{equation}\label{R1+R2-TM}
G^{\rm (TM)}_{(1+2)}(\tau)= 360
\tau^5(e^{\Omega}-1)\int_0^{\infty}d\Omega' \int_0^{\Omega'} dK\,
\frac{K\,\left[K^2-\Omega'\left(\Omega+\Omega'\right)\right]^2} 
{\sqrt{(\Omega'^2-K^2)\left[(\Omega+\Omega')^2-K^2\right]}}
\,
\frac{e^{\Omega'}}{\left(e^{\Omega'}-1\right)\left(e^{\Omega+\Omega'}-1\right)}, 
\end{equation}
whereas the contribution from $R_3$ gives
\begin{equation}\label{R3-TM}
G^{\rm (TM)}_{(3)}(\tau)= 360
\tau^5\int_0^{\Omega/2}d\Omega' \int_0^{\Omega'} dK\,
\frac{K\,\left[K^2+\Omega'\left(\Omega-\Omega'\right)\right]^2} 
{\sqrt{(\Omega'^2-K^2)\left[(\Omega-\Omega')^2-K^2\right]}}
\,\left(
\frac{1}{e^{\Omega'}-1}+
\frac{1}{e^{\Omega-\Omega'}-1}\right).
\end{equation}

In the high temperature limit, the contribution  
from Region $R_3$ is ${\cal O}(\tau)$ and hence  
negligible as in the scalar case, whereas 
the contribution from $R_1$ and $R_2$ is derived from (\ref{R1+R2-TM}) by 
neglecting $\Omega$ in the integrand and by taking  $e^{\Omega}-1\approx \Omega.$
We find $G^{\rm (TM)}(\tau\rightarrow\infty)=G^{\rm (TE)}(\tau\rightarrow\infty)[=
G^{\rm D}(\tau\rightarrow\infty)].$

For low temperatures, we proceed as in the discussion of the scalar case: the contributions from 
$R_1$ and $R_3$ are identical, and from $R_2$ negligible. We find 
$G^{\rm (TM)}(\tau\rightarrow 0)= 48\pi^4\tau^4.$ When added to 
$G^{\rm (TE)}(\tau\rightarrow 0)= 16\pi^4\tau^4,$ we get the electromagnetic 
result as discussed in Sec.~III.

\section{Field free energy with a static mirror}

We discuss both scalar (with Dirichlet boundary conditions) and 
electromagnetic field models. 
We first consider two parallel mirrors at rest separated by a distance $L,$ and 
compute the free energy in the volume $V=AL$ contained within the mirrors. 
Then, we take the limit $L\rightarrow\infty$ so as to identify the single-mirror effect.
In this limit, we find a term proportional to $V,$ corresponding to the free-space case, 
and, in the scalar case, a term  proportional to $A,$ which does not depend on $L.$  The latter contains the independent 
single-mirror effects of the two mirrors, with only one side of each mirror taken into
account. By symmetry, this is equal to the effect of one mirror when  both sides
are taken into account. 
The free energy for two parallel plates has been calculated  by 
several different methods~\cite{free_temp}. Here we follow the approach of 
Ref.~\cite{plunien}.

The partition function is defined as 
\begin{equation}\label{Z}
Z = {\rm tr}\, \exp[-H_0/(k_B T)],
\end{equation}
with $H_0=\sum_{\lambda} \hbar \omega_{\lambda}(a_{\lambda}^{\dagger}a_{\lambda}+1/2).$ 
$\lambda\equiv\{n,{\bf k}_{\parallel}\}$ represents a set of parameters  defining a given field mode
(which also contains an index for polarization in the electromagnetic case),
with 
$\omega_{\lambda}=\sqrt{(n\pi/L)^2+k_{\parallel}^2}.$ Eq.~(\ref{Z}) yields
\begin{equation}
Z = \Pi_{\lambda}\frac{e^{-\hbar\omega_{\lambda}/(2k_B T)}}{1-e^{-\hbar\omega_{\lambda}/(k_B T)}},
\end{equation}
which combined with (\ref{def_F}) leads to
\begin{equation}\label{Ffirst}
{\cal F}=\sum_{\lambda}\frac{\hbar\omega_{\lambda}}{2}+k_B T \sum_{\lambda}
\ln\left(1-e^{-\hbar\omega_{\lambda}/(k_B T)}\right).
\end{equation}
The first term in (\ref{Ffirst}) represents the zero point energy ${\cal E}_{\rm vac}.$ 
We want to compute the second term, which contains the 
temperature dependence:
\begin{equation}\label{Fsum}
{\cal F}_T = \sum_{n=0}^{\infty} g(n) a(n),
\end{equation}
where
\begin{equation}\label{Fsecond}
a(n)\equiv
\frac{A k_B T}{2\pi} \int_0^{\infty} d k_{\parallel} k_{\parallel}
\ln\left[1-\exp\left(-\frac{\hbar\sqrt{(n\pi/L)^2+k_{\parallel}^2}}{k_B T}\right)\right],
\end{equation}
with $g(n)=1-\delta_{n0}$ and $g(n)=2-\delta_{n0}$ in the scalar and  electromagnetic models, respectively.
Hence in the former case grazing modes,  corresponding to $n=0,$ are ruled out, whereas
in the electromagnetic model TM polarized grazing modes provide a non vanishing contribution 
to the free energy. 

In order to compute ${\cal F}_T,$ we employ Poisson sum formula
\begin{equation}\label{Poisson}
\frac{1}{2}a(0)+\sum_{n=1}^{\infty}a(n)=A_0+2\sum_{j=1}^{\infty}{\rm Re}A_j,
\end{equation}
where 
\begin{equation}
A_j=\int_0^{\infty}dn\, a(n) e^{i 2 j \pi n}.
\end{equation}
We change the variable of integration from 
$k_{\parallel}$ to $\xi\equiv \hbar \sqrt{(n\pi/L)^2+k_{\parallel}^2}/(k_B T),$ and integrate 
over $n$ to find
\begin{equation}\label{A0}
A_0 = -\frac{\pi^2}{90} \frac{V (k_B T)^4}{\hbar^3},
\end{equation}
and for non vanishing values of $j,$ 
\begin{equation}\label{Aj}
A_j=\frac{A (k_B T)^3}{2\pi^2 j \hbar^2}\int_0^{\infty} d\xi \xi \ln(1-e^{-\xi})
\frac{e^{i \Lambda_j \xi}-1}{2 i},
\end{equation}
where $\Lambda_j = 2 j L k_B T/\hbar.$
By integrating the r.-h.-s. of (\ref{Aj}) by parts, we find
\begin{equation}\label{Aj2}
{\rm Re}A_j=\frac{A (k_B T)^2}{8\pi^2 j^2 \hbar L}
\left\{-\frac{1}{2\Lambda_j^2}\left[\pi\Lambda_j\coth(\pi\Lambda_j)-1\right]
+{\rm Re}
\int_0^{\infty} d\xi\, \frac{\xi}{e^{\xi}-1}
e^{i \Lambda_j \xi}\right\}.
\end{equation}
We calculate the r.-h.-s. of (\ref{Aj2}) in the limit $\Lambda_j\gg 1.$
By successive integrations by parts, one may show that the last term within 
brackets in (\ref{Aj2}) yields $-1/(2\Lambda_j^2)+{\cal O}(1/\Lambda_j^3)$
and may be neglected in (\ref{Aj2}):
\begin{equation}
{\rm Re}A_j=-\frac{A k_B T}{32\pi j^3 L^2}
[1 + {\cal O}(1/\Lambda_j)].
\end{equation}
Hence in the limit $L\rightarrow\infty$ 
only $A_0$ contributes in the Poisson sum formula Eq.~(\ref{Poisson}). 

Finally, when computing the free energy for the scalar Dirichlet field,
we need to subtract $a(0)/2$
from the r.-h.-s. of (\ref{Poisson}), since 
$g(0)=0$ in (\ref{Fsum}) in this case.
We calculate $a(0)$ from (\ref{Fsecond}):
\begin{equation}\label{Fgrazing}
a(0)= -\frac{\zeta(3)}{2\pi}\frac{A (k_B T)^3}{\hbar^3}.
\end{equation}
Combining (\ref{Fsum}), (\ref{Poisson}), (\ref{A0}) and (\ref{Fgrazing}), we derive 
the free energy for the Dirichlet field and a single static plate as given by 
(\ref{F_scalar}). For the electromagnetic field, on the other hand, we may employ 
Poisson formula  directly, since in this case the sum over $n$ in 
(\ref{Fsum}) has  already  the form of the l.-h.-s. of  (\ref{Poisson}). Then we find
 ${\cal F}_T^{\rm EM}=2 A_0$ [with $A_0$ given by 
(\ref{A0})], in agreement with (\ref{F_em}).

\section{Thermal energy density for the Dirichlet scalar field with a static mirror}

In this appendix, we compute the energy density 
\begin{equation}\label{energy_density}
w(z)={1\over 2} \langle (\partial_t\phi)^2+(\nabla\phi)^2 \rangle
\end{equation} 
for a scalar field 
at temperature $T,$ and with Dirichlet boundary conditions at $z=0.$ 
Since we consider the static case, we replace $\phi$ by $\phi^0,$ 
and from its normal mode expansion as given by Eq.~(\ref{phi0escalar})
derive (with $k=\sqrt{k_{\|}^2+k_z^2}\,$)
\begin{equation}\label{dens_parcial}
w(z)={\hbar\over 2 \pi^2}\int_0^{\infty}dk_z\int_0^{\infty} dk_{\|} {k_{\|}\over k}
\left[\left(k^2+k_{\|}^2\right)\sin^2(k_z z) +k_z^2 \cos^2(k_z z)\right]
\left[{1\over 2}+{\overline n}(k)\right].
\end{equation}
As in the derivation of the susceptibility $\chi(\omega),$ the energy density
is naturally split into two contributions, one from vacuum fluctuations [the `$1/2$' 
inside brackets in (\ref{dens_parcial})], the other from thermal fluctuations, 
which corresponds to the factor ${\overline n}(k)$ in (\ref{dens_parcial}).
Accordingly, we write $w=w_T+w_{\rm vac}.$ Here we are only interested in the thermal 
contribution $w_T,$ which is obtained from 
(\ref{dens_parcial})
after changing the variable of integration from $k_{\|}$ to $k:$
\begin{equation}\label{dens_1}
w_T(z)={\pi^2\over 30} {(k_B T)^4\over \hbar^3}+\Delta w_T.
\end{equation}
The first term in (\ref{dens_1}) represents the free-space energy density for a scalar field; 
the effect of the boundary at $z=0$ is contained in $\Delta w_T:$
\begin{equation}\label{res_energy_dens}
\Delta w_T(z) = {\hbar \over 8 \pi^2}\left[{1\over 2 z^4}-{\pi\over 2} {k_B   T\over \hbar z^3}
\coth\left({2\pi k_B T z\over \hbar}\right) -\pi^2 \left({k_B T\over \hbar z}\right)^2
{1\over \sinh^2\left({2\pi k_B T z\over \hbar}\right)}   \right]. 
\end{equation}
$\Delta w_T(z)$ is a negative-defined, increasing function of $z$ that 
goes to zero as $-kT/(16\pi z^3)$ for $z\gg \hbar/(k_B T).$ Hence the mirror 
reduces the thermal energy density, an effect stronger near the mirror. $\Delta w_T$ is
finite at $z=0,$ and vanishes at $T=0$ as expected from its definition. 

The total modification of the internal energy is given by the volume integral of 
$\Delta w_T(z),$ taking into account both 
sides of the mirror:
\begin{equation}
\Delta U_T= 2 A \int_0^{\infty} dz\, \Delta w_T(z)= {\pi\over 4} A {(k_B T)^3\over \hbar^2}
\int_{-\infty}^{\infty} d{\cal Z} \left[ {2\over {\cal Z}^4}-{\coth{\cal Z}\over {\cal Z}^3}
-{1\over {\cal Z}^2\sinh^2{\cal Z}}\right].
\end{equation}
The integral above may be computed with the method of residues, by taking a semi-circular 
path  (radius $\rightarrow\infty$) in the complex plane of ${\cal Z}.$ The poles of the integrand lie
along the imaginary axis, at the positions ${\cal Z}= i n\pi,$ with $n$ integer, $n\neq 0.$ We find
\begin{equation}\label{energia_final}
\Delta U_T= - {1\over 2\pi} A {(k_B T)^3\over \hbar^2}\sum_{n=1}^{\infty}{1\over n^3}.
\end{equation}
The series appearing in (\ref{energia_final}) is equal to 
$\zeta(3).$  Then, comparing (\ref{res_mass}) with (\ref{energia_final}), we find 
$\Delta U_T=\Delta m \, c^2$ as discussed in Sec.~V.


\begin{thebibliography}{99}
%

\bibitem{Casimir} H. B. G. Casimir, Proc. K. Ned. Akad.\ Wet. \textbf{51},
793\textbf{ }(1948).

\bibitem{barton91} G. Barton, J. Phys. (London) A: Math. 
Gen. {\bf 24}, 5533 (1991); C. Eberlein, {\it ibid.} {\bf 25}, 3015 (1992).

\bibitem{jrqoptics} V. B. Braginsky and F. Ya. Khalili,
Phys. Lett. A {\bf 161}, 197 (1991); 
M. T. Jaekel and S. Reynaud, Quantum Opt. {\bf 4}, 39 (1992).
%

\bibitem{Moore} G. T. Moore, J. Math. Phys. \textbf{11}, 2679 (1970).
%

\bibitem{FDavies} S. A. Fulling and P. C. W. Davies, Proc. R. Soc. London A {\bf 348}, 393 (1976).
%

\bibitem{Ford} L. H. Ford  and A. Vilenkin, Phys. Rev. D \textbf{25,} 2569
(1982).

\bibitem{PA94} P. A. Maia Neto, J. Phys. A \textbf{27}, 2167 (1994).

\bibitem{PRA93} G. Barton, {\it New aspects of the Casimir effect: fluctuations and 
radiative reaction,} in {\it Cavity Quantum Electrodynamics,} Supplement: Advances in 
Atomic, Molecular and Optical Physics, edited by
P. Berman (Academic Press, New York, 1993);
P. A. Maia Neto and S. Reynaud, Phys. Rev. A {\bf 47}, 1639 (1993).


\bibitem{JaekelReynaudPLA93} M. T. Jaekel and S. Reynaud, Phys. Lett. A {\bf 172}, 319 (1993).

\bibitem{thermal} A. Lambrecht, M. T. Jaekel and S. Reynaud, Europhys. Lett. {\bf 43},  147 (1998);
 G. Plunien, R. Sch\"utzhold and G. Soff, Phys. Rev. Lett. {\bf 84}, 1882 (2000);
J. Hui, S. Qing-Yun and W. Jian-Sheng, Phys. Lett. A {\bf 268}, 174 (2000);
R. Sch\"utzhold, G. Plunien and G. Soff, Phys. Rev. A {\bf 65}, 043820 (2002).


\bibitem{kubo} H. B. Callen and T. A. Welton,  Phys. Rev. {\bf 83},
34 (1951); R. Kubo, Rep. Prog. Phys. {\bf 29}, 255 (1966). 
 

\bibitem{bogo} V. V. Dodonov, A. B. Klimov and V. I. Man'ko, Phys. Lett. A {\bf 149}, 225 (1990); 
P. A. Maia Neto and L. A. S. Machado, Brazilian J. Phys.
\textbf{25}, 324 (1995).

\bibitem{bogo2} P. A. Maia Neto and L. A. S. Machado, Phys. Rev. A {\bf 54}, 3420 (1996).

\bibitem{inertia-one-plate} G. Barton and C. Eberlein, 
Ann. Phys. (N.Y.) {\bf 227}, 222 (1993). 

\bibitem{eberlein}  R. G\"{u}tig and C. Eberlein, J. Phys. (London) A: Math. 
Gen. {\bf 31}, 6819 (1998).


\bibitem{barton-calogeracos}  G. Barton and A. Calogeracos, Ann. Phys. (NY) {\bf 238},
227 (1995); A. Calogeracos and 
G. Barton, {\it ibid.} {\bf 238}, 268
(1995).



\bibitem{abramowitz} M. Abramowitz and I. Stegun, {\it Handbook of Mathematical Functions} (Dover, New York, 1972).


\bibitem{PRA98} D. F. Mundarain and P. A. Maia Neto, Phys. Rev. A {\bf 57}, 1379 (1998). Appendix A
presents a detailed derivation of the 
boundary conditions.


\bibitem{PRL2000} D. A. R. Dalvit and P. A. Maia Neto, Phys. Rev. Lett. {\bf 84}, 798 (2000).

\bibitem{decoherence} P. A. Maia Neto and  D. A. R. Dalvit, Phys. 
Rev. A {\bf 62}, 042103 (2000).

\bibitem{lambrecht_prl} A. Lambrecht, M. T. Jaekel and S. Reynaud, Phys. Rev. Lett. {\bf 77}, 615 (1996).

\bibitem{foot_2} 
For times much larger than $2\pi/\omega_0,$ 
the diffusion coefficient $D(t)$ tends to an 
asymptotic constant value, which 
is related to the spectrum of 
fluctuations of the field taken at the particular frequency $\omega_0.$ 
This allows us to employ the fluctuation-dissipation theorem and derive 
a simple relation between  
$D(t\gg 2\pi/\omega_0)$ and the damping rate $\Gamma,$ which results in 
(\ref{descoerencia-dissipacao}). For consistency of the derivation, the decoherence time scale 
must be larger than $2\pi/\omega_0.$ 
Thus, as in the discussion of damping, we are not allowed to consider the 
free particle limit ($\omega_0=0$). 
The limit of `fast' decoherence (i.e., faster 
than the free evolution time scale)
in the high temperature approximation 
 was considered by Maia Neto and 
Dalvit [{\it Decoherence effects of motion-induced radiation}, in 
{\it Modern challenges in quantum optics,} edited by M. Orszag and J. C. Retamal
(Springer, Berlin, 2001)]. 
In this case, 
the mirror does not have time to probe the potential well 
before decaying into an incoherent statistical mixture, and its 
effect is negligible as regards decoherence. The 
free particle case in the high temperature limit is also 
discussed in Ref.~\cite{joos}. 

\bibitem{joos} E. Joos and H. D. Zeh, Z. Phys. B {\bf 59}, 223 (1985).

\bibitem{deco_ref} W. H. Zurek, S. Habib and J. P. Paz, Phys. Rev. Lett. {\bf 70}, 1187 (1993).

\bibitem{braginsky} V. B. Braginsky and A. B. Manukin, Sov. Phys. JETP {\bf 25}, 653 (1967);
A. B. Matsko, E. A. Zubova and S. P. Vyatchanin, Optics Comm. {\bf 131}, 107 (1996). 

\bibitem{JRmass} M.-T. Jaekel and S. Reynaud, J. Phys. I (France) {\bf 3}, 1093 (1993).

\bibitem{Ddimensions}  L. A. S. Machado and P. A. Maia Neto, Phys. Rev. D {\bf 65}, 125005 (2002).

\bibitem{kardar} R. Golestanian and M. Kardar, Phys. Rev. Lett. {\bf 78}, 3421 (1997); Phys. Rev. A 
{\bf 58}, 1713 (1998).

\bibitem{brown-maclay} L. S. Brown and G. J. Maclay, Phys. Rev. {\bf 184}, 1272 (1969).

\bibitem{foot_3} Remarkably, the mass dependence in $\Gamma$ [see (\ref{gamma_1})] is canceled by 
the mass introduced by $\Delta Z_0$ in (\ref{descoerencia-dissipacao}), so that 
$\Gamma^{\rm dec}$ does not depend on the mass of the mirror, for any $T.$ 
Moreover, both $\Gamma$ and $\Gamma^{\rm dec}$
do not depend on $\omega_0$ in the high-temperature limit, provided that $\omega_0\gg \Gamma, \Gamma^{\rm dec}$
(see~\cite{foot_2}).  

\bibitem{free_temp} 
M. Fierz, Helv. Phys. Acta {\bf 33}, 855 (1960);
J. Mehra, Physica {\bf 37}, 145 (1967);
 J. Ambjorn and S. Wolfram, 
Ann. Phys. NY {\bf 147}, 1 (1983);
N. F. Svaiter, Nuovo Cimento A {\bf 105}, 959 (1992); M. V. Cougo-Pinto, C. Farina and 
A. Tort, Lett. Math. Phys. {\bf 37}, 159 (1996). 

\bibitem{plunien} G. Plunien, B. M\"uller and W. Greiner, Phys. Rep. {\bf 134}, 88 (1986). 

\end{thebibliography}
\end{document}